# Mathematical Model of Emotional Habituation to Novelty
# -Modeling with Bayesian Update and Information Theory-*

Takahiro Sekoguchi, Yuki Sakai, and Hideyoshi Yanagisawa

*Abstract*— Novelty is an important factor of creativity in product design. Acceptance of novelty, however, depends on one's emotions. Yanagisawa, the last author, and his colleagues previously developed a mathematical model of emotional dimensions associated with novelty such as arousal (surprise) and valence (positivity and negativity). The model formalized arousal as Bayesian information gain and valence as a function of arousal based on Berlyne's arousal potential theory. One becomes accustomed to novelty by repeated exposure. This so-called habituation to novelty is important in the design of long-term product experience. We herein propose a mathematical model of habituation to novelty based on the emotional dimension model. We formalized the habituation as a decrement in information gain from a novel event through Bayesian update. We derived the information gained from the repeated exposure of a novel stimulus as a function of three parameters: initial prediction error, initial uncertainty, and noise of sensory stimulus. With the proposed model, we discovered an interaction effect of the initial prediction error and initial uncertainty on habituation. Furthermore, we demonstrate that a range of positive emotions on prediction errors shift toward becoming more novel by repeated exposure. We hypothesize that the ease to become accustomed to novelty depends on the initial uncertainty. To verify this hypothesis, we conducted an experiment with several short videos in which different percussion instruments were played. We manipulated the uncertainty level by the popularity of instruments and prediction error by the congruity between sounds and videos. We used event-related potential P300 amplitudes and the subjective reports of surprise in response to the sounds as measures of arousal levels. The experimental results supported our hypothesis; therefore, the decrement in information gain can be decomposed into initial prediction error and initial uncertainty, and is considered as a valid measure of emotional habituation.

## I. Introduction

Novelty is an important factor of creativity in product design. However, novelty acceptance differs by individual. Raymond Loewy, a pioneer of industrial design, defined a broader term between novelty attraction and fear of the unknown as MAYA (most advanced, yet acceptable), which is important for new designs to be widely accepted in society [1]. Berlyne advocated the theory that a range exists that maximizes the pleasant feelings between familiar and novel, and between simple and complex [2]. Silvia et al. validated Berlyne's hypothesis through experiments with shapes and works of art [3]. They revealed that emotions for novelty differed according to characteristics such as an individual's knowledge. Yanagisawa, the last author, and his colleagues developed a mathematical model of emotional dimensions associated with novelty that considered personal characteristics using information theory and the Bayes model [4]. The model formalized arousal (primary emotional dimension) as a function of prediction error, uncertainty, and external noise. The function model was supported experimentally using the event-related potential (ERP) P300 of human participants as an index of arousal. Furthermore, they formalized valence as a function of arousal based on Berlyne's arousal potential theory.

One becomes accustomed to novelty through its repeated exposure. Psychologists define the response extinction to repeated presentation of stimuli as habituation [5]. Habituation to novelty is important in the design of long-term product experience. Croy et al. suggested that repeated presentation would result in the reduced emotional salience of unpleasant stimuli from their experimental results using ERP [6]. Lévy et al. demonstrated that the range of novelty that maximizes pleasant feelings shifts to a more novel and complicated direction by repeated stimuli [7]. However, mathematical analyses were not performed in these studies, and biases from factors such as one's prior knowledge and experience were not investigated exhaustively.

In this study, we aim at developing a general mathematical model representing emotional habituation. We considered it useful to elucidate habituation on novelty and modeling its laws mathematically to design creative products acceptable to society. We applied and developed the model of Yanagisawa and described the framework of a mathematical habituation model using information theory. We subsequently considered the effects of the initial prediction error and initial uncertainty on the time-series change of emotion as the primary factors constituting novelty by simulation of the model. Finally, we conducted experiments and compared the results with the simulation results of this study for validation.

*Research supported by JSPS KAKEN grant number 18H03318.
T. Sekoguchi is a graduate student of Department of Mechanical Engineering, The University of Tokyo, Japan.
Y. Sakai is a research student in Department of Mechanical Engineering, The University of Tokyo, Japan.
Dr. H. Yanagisawa is co-director of Design Engineering Laboratory and associate professor of Department of Mechanical Engineering, The University of Tokyo, Japan (phone: +81-3-5841-6330; e-mail: hide@mech.t.u-tokyo.ac.jp).

## II. MODELING HABITUATION TO NOVELTY

### A. Modeling Based on Information Theory

In this section, we describe a method for modeling habituation mathematically. Because this method is based on the model of Yanagisawa, we explain the essential parts briefly and focus on the original parts.

We defined novelty as the amount of information a person acquires by experiencing an event. The information content given by an event (self-information) is consistent with its uncertainty before experiencing it. An expected value of self-information (information entropy) of the prior distribution represents the uncertainty of the prior expectation. An information entropy of the posterior distribution represents the uncertainty of the posterior experience by an event. Considering a transition before and after experiencing an event, the information entropy decreases by experiencing an event. This decrease is proportional to the information acquired from the event, which is called the *information gain*. An event with large novelty yields a large information gain and causes salient emotions such as surprise [8]. Generally, emotions are arranged spatially in two dimensions primarily: arousal and valence. Because surprise is positioned as an emotional state with high arousal, information gain is considered to correspond to arousal.

We expressed habituation using information theory. Repeating the same stimulus decreases the information gain, as the prior distribution approaches the posterior distribution gradually. Therefore, we consider habituation as a decrement in information gain. We describe the modeling method of habituation mathematically in the following section.

### B. Bayesian Update Model

Let a prior be $\pi(\theta)$ in terms of a parameter $\theta$ that one estimates. After one obtains continuous data $x \in R$ by experiencing an event, the prior $\pi(\theta)$ is updated to the posterior $\pi(\theta|x)$ according to the following equation derived from Bayes's theorem:

$$\pi(\theta|x) = \frac{f(x|\theta)^\alpha \pi(\theta)}{\int_\theta f(x|\theta)^\alpha \pi(\theta) d\theta} \propto f(x|\theta)^\alpha \pi(\theta), \tag{1}$$

where $f(x|\theta)$ is a likelihood function of $\theta$ when data $x$ is obtained. $\alpha$ is a parameter that adjusts the amount of change of the information gain, which is called the *learning rate* [9].

We consider the posterior when experiencing events $k$ times, $\pi_k(\theta|x)$ to be the prior when experiencing events $k+1$ times, $\pi_{k+1}(\theta)$. Assuming that the likelihood functions are independent simultaneous distributions, the order of data collection is irrelevant. Thus, the posterior $\pi_k(\theta|x)$ is obtained from the following equation:

$$\pi_k(\theta|x) \propto \prod_{i=1}^{k} f(x_i|\theta)^\alpha \pi(\theta). \tag{2}$$

$\pi_k(\theta|x)$ is proportional to the product of each likelihood function and the initial prior.

Assume one obtains $n$ samples of events $x$ and encodes them as a Gaussian distribution $N(\mu, \sigma^2)$ with a flat prior. Further, assume a nonflat prior of $\mu$ that follows a Gaussian distribution $N(\eta, \tau^2)$. Using Bayes's theorem, the prior when experiencing an event $n$ times is updated to a Gaussian distribution $N(\eta_n, \tau_n^2)$, where

$$\eta_n = \frac{\alpha n S_{pI} \bar{x} + S_l \eta}{\alpha n S_{pI} + S_l}; \quad \tau_n^2 = \frac{S_{pI} S_l}{\alpha n S_{pI} + S_l}. \tag{3}$$

In these equations, $\bar{x}$ is the mean of the data, $S_{pI} = \tau^2$, and $S_l = \sigma^2$.

### C. Arousal Update Model (Habituation)

The information gain from the prior to the posterior $G$ can be derived from (3) as follows

$$G = \int_{-\infty}^{\infty} \pi(\mu|x) \ln \frac{\pi(\mu)}{\pi(\mu|x)} d\mu = \frac{1}{2}(A + B\delta_I^2), \tag{4}$$

$$A = \frac{g_{n-1}}{g_n} - \ln \frac{g_{n-1}}{g_n} - 1, \quad B = \frac{\alpha^2 S_{pI} S_l}{g_{n-1} g_n^2},$$

$$g_x = \alpha S_{pI} x + S_l,$$

where $\delta_I = |\eta - \bar{x}|$ is the difference between the prior expectation $\eta$ and peak of the likelihood function $\bar{x}$. It represents the difference between expectations and reality. Thus, we termed $\delta_I$ the *initial prediction error*. The information entropy of the prior is proportional to the logarithm of $\tau^2$. Thus, we termed $S_{pI}$ the *initial uncertainty*. $S_l = \sigma^2$ represents the variance of the data. In the case of sensory data (i.e., stimuli), the variance refers to the *external noise*. From (4), we can regard the information gain $G$ as a function of the initial prediction error $\delta_I$, initial uncertainty $S_{pI}$, and external noise $S_l$.

*D. Formalization of Acceptable Prediction Error*

Berlyne assumed that a range maximizing the pleasure between the novel and familiar, and between simple and complex stimuli exists. Further, he assumed that these hedonic qualities of stimuli arise from separate biological incentivization systems: the *reward system* and the *aversion system*, each of which is represented by a sigmoid function. The joint operation of these two systems create an inverted U-shaped curve called Wundt curve. The valence of a stimulus changes from neutral to positive as the arousal increases but shifts from positive to negative once the arousal passes the peak positive valence. The range of pleasant feelings is obtained from the following equation:

$$\delta_g^2 = \frac{1}{B}\left|\frac{1}{c}\ln\frac{h_a e^{G_r} - h_r e^{G_a}}{h_r - h_a} - A\right| . \quad (5)$$

In (5), $G_r$, $h_r$, and $c$ represent the thresholds of the information gain that activate the reward systems, the maxima of positive valence levels, and the gradients, respectively. $G_a$ and $h_a$ represent these in aversion systems.

III. Effect of Initial Prediction Errors and Initial Uncertainties on Emotional Dimensions

*A. Habituation (Update of Arousal)*

We first analyzed how initial prediction errors, initial uncertainties, and external noise affect the updates of prediction errors and uncertainties. From (3), the partial derivatives of the prediction error and the uncertainty with respect to updating times $n$ are the following:

$$\frac{\partial \delta_n}{\partial n} = -\frac{\alpha \delta_I S_{pI}/S_l}{(\alpha n S_{pI}/S_l + 1)^2} , \quad (6)$$

$$\frac{\partial S_{pn}}{\partial n} = -\frac{\alpha/S_l}{(\alpha n/S_l + 1/S_{pI})^2} , \quad (7)$$

where $\delta_n$ and $S_{pn}$ are the prediction error and uncertainty when experiencing events $n$ times, respectively. Equation (6) implies that the prediction error decreases as $n$ increases, and the decay rate of prediction error increases with the ratio of the initial uncertainty to noise ($S_{pI}/S_l$). Meanwhile, (7) implies that uncertainty decreases as $n$ increases, and the decay rate of uncertainty decreases as the initial uncertainty $S_{pI}$ or the noise $S_l$ increases.

We subsequently analyzed how the initial prediction error and initial uncertainty affect the updates of information gain. Fig. 1 shows the updates of information gain when the initial uncertainty is fixed. At any $n$, the information gain increases with the initial prediction error. The decay rate of information gain increases with the initial prediction error because the partial derivative of the decay rate of information gain with respect to the initial prediction error is always less than zero:

$$\frac{\partial}{\partial \delta_I}\left(\frac{\partial G}{\partial n}\right) < 0 \ (\delta_I > 0). \quad (8)$$

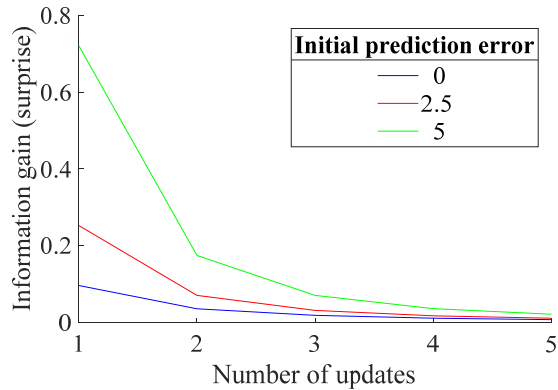

Figure 1. Updates of information gain for different initial prediction errors (initial uncertainty = 1.0, noise = 0.5, $\alpha$ = 0.1)

Figs. 2 and 3 show the updates of information gain when the initial prediction errors are fixed in two cases. In the case of $n = 1$, larger initial uncertainties result in larger information gains when the initial prediction error is 4. Meanwhile, when the initial prediction error is 10, larger initial uncertainties result in smaller information gains. That is, the two functions of different uncertainties exhibit an intersection and the magnitude relation of the information gains changes as the initial prediction error increases when number of updates is 1. We found that this condition applies when the relationship between different initial uncertainties $S_{p1}$ and $S_{p2}$ is as follows:

$$S_{p1} \cdot S_{p2} > \left(\frac{S_l}{\alpha}\right)^2. \tag{9}$$

As the number of updates increases, the difference in information gains between the two functions changes, as well. When the initial prediction error is 4, the magnitude relation of the information gains with different uncertainties is reversed by updating. When the initial prediction error is 10, the information gain with a larger initial uncertainty is always greater. Furthermore, the information gain with a larger initial uncertainty decreases more significantly from $n = 1$ to $n = 2$ and converges to zero faster at both of the initial prediction errors. These imply that a lower initial uncertainty tends to result in a larger information gain by repeating the stimulus.

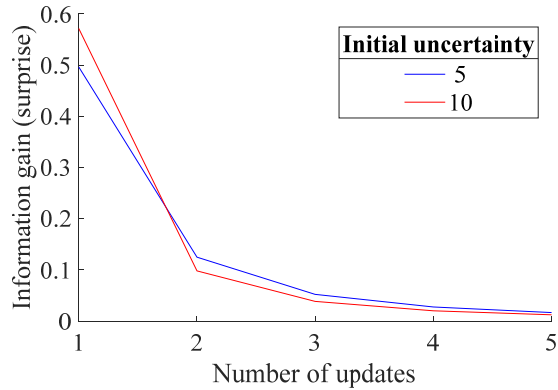

Figure 2. Updates of information gain for different initial uncertainties (initial prediction error = 4.0, noise = 0.5, $\alpha$ = 0.1)

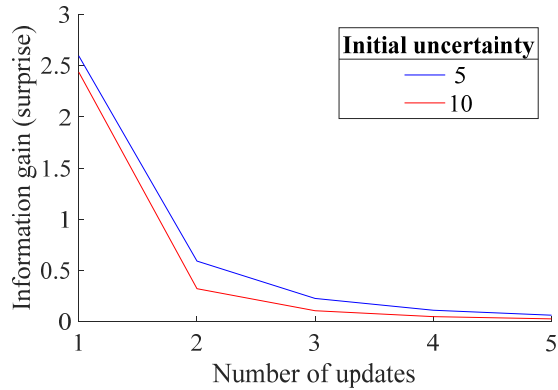

Figure 3. Updates of information gain for different initial uncertainties (initial prediction error = 10, noise = 0.5, $\alpha$ = 0.1)

### B. Updates of Acceptable Prediction Error

We investigated how the initial uncertainty affects the updates of the acceptable prediction error range, as shown in (5), by updating. Fig. 4 shows the initial prediction error increase by updating, where the valence shifts from positive to negative. This is proven mathematically because the partial derivative of the acceptable range with respect to number of updates is always more than zero:

$$\frac{\partial \delta_g^2}{\partial n} > 0 \; (S_{pI} > 0). \tag{10}$$

Fig. 4 also shows the larger initial uncertainties result in a larger increase rate of the acceptable prediction error range.

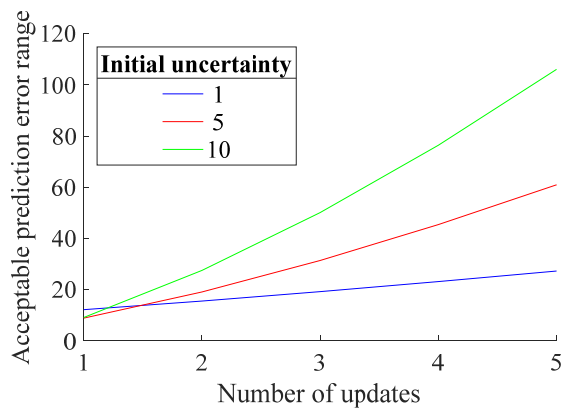

Figure 4. Updates of acceptable prediction error range for different initial uncertainties (noise = 0.5, $\alpha = 0.1$)

*C. Meaning of Simulation Results*

We demonstrated an interaction effect between the initial prediction error and initial uncertainty on the decrement in information gain. The larger the initial uncertainty, the sooner the information gain decreases and converges to zero. The total value of the information gain from $n = 1$ to $n = 5$ increases with the uncertainty when the initial prediction error is 4, but decreases when the initial prediction error is 10. This implies that one is surprised for a long period even if the uncertainty of the event is not particularly large. This phenomenon tends to occur as the initial prediction error becomes larger. As Silvia et al. demonstrated in their study, we assume that the uncertainty of an event depends on the knowledge and experience that the person possesses, the attribute of the object, such as the affinity that comes from the typicality of the object, and the relationship between the person and object. This result can be useful for designing products that one can enjoy for a long period. Designers can maximize the surprise experienced while using the product by designing its appearance, even if it is familiar.

Furthermore, we demonstrated that the positive range of the Wundt curve shifted to a more novel direction by updating. This implies that people, by repeating stimuli, prefer more novel and complicated objects gradually. The experimental results of [7] support the validity of this result. Furthermore, we demonstrated that the increase rate of acceptable prediction error range becomes larger as the initial uncertainty increases. This implies that an event with high uncertainty tends to change the acceptable prediction error range less dynamically. This result matches with our intuition. For example, children tend to be interested in things (e.g., how to play instruments or how to speak a foreign language) earlier than adults. We suggest that the uncertainty of an event can explain the difference in the experience of the event between children and adults.

## IV. EXPERIMENT

We investigated the effects of uncertainty and prediction errors on emotional habituation to validate our hypothesis derived from the mathematical model. Specifically, we focused on how the initial uncertainty affects the decrement in surprise. The procedure of this experiment is based on the method of [4]. A set of short videos featuring percussion instruments and accompanying sounds were used as stimuli. In each video, a percussion instrument was presented and subsequently beaten. Different percussive sounds were synthesized. We set a transition from a visual prior (i.e., the appearance of an instrument) to an auditory posterior (i.e., the percussive sound). Participants predicted an instrument's sound from its appearance and subsequently listened to a sound. We induced prediction errors by manipulating the congruency between the synthesized percussive sounds and the instrument shown. We assumed that the prediction errors were large when the synthesized percussive sounds were incongruent with the instruments shown, and we assumed that the familiarity or unfamiliarity of the instruments shown produced different levels of uncertainty. The appearance of a familiar percussion instrument, such as a hand drum, produces the certainty in expectations concerning its sound (i.e., a small uncertainty). The appearance of an unfamiliar percussion instrument, such as the African percussion instrument known as the jawbone, produces uncertain expectations concerning its sound (i.e., a large uncertainty).

We used both questionnaires and ERP recordings to assess the participants' levels of surprise in response to the percussive sound in each video. We quantified surprise intensities based on responses to a four-level Likert scale and measurements of ERP P300 amplitudes [10].

*A. Method*

Eight right-handed healthy male volunteers (range: 21–27 years) with normal or corrected-to-normal vision and hearing participated in this study. The study protocol was approved by the Ethics Committee of the Graduate School of Engineering at the University of Tokyo. In accordance with the principles of the Declaration of Helsinki, all participants provided written informed consent prior to their participation in this study. The participants were allowed to stop the experiment sessions at their convenience.

The stimuli consisted of eight short videos in which a percussion instrument was beaten once and followed by a synthesized percussive sound. Table 1 shows the combinations of instruments shown and the synthesized sounds. The clave and hand drum were selected as familiar percussion instruments (type A), and the jawbone and slit drum were selected as unfamiliar percussion instruments (type B). To create incongruent conditions, we synthesized percussive sounds that were inconsistent with the instruments shown. Our stimuli included videos with visually familiar instruments and congruent sounds (type AX), videos with visually familiar instruments and incongruent sounds (type AY), videos with visually unfamiliar instruments and congruent sounds (type BX), and videos with visually unfamiliar instruments and incongruent sounds (type BY).

The duration of each video was 2,500 ms. First, a percussion instrument appeared in the center of the screen. The percussion instrument was subsequently beaten once 500 ms into the video while a percussive sound was presented simultaneously. Each video exhibited an 18° horizontal visual angle and a 10° vertical visual angle, and was presented centrally against a black background on a 29.8-inch display located 100 cm from the participant. The participants wore noise-cancelling headphones covered by earmuffs while watching the videos.

The participants completed the experiments individually in an electromagnetically shielded dark room. After the participants received instructions for the procedure, they were asked to start the experiment.

First, we conducted sound-only experiments in which we attempted to ensure uniform surprise levels in response to the percussive sounds used in each video type (i.e., AX, AY, BX, and BY). Achieving this uniformity was necessary such that we could be sure that our observations in later experiments with audiovisual stimuli reflected the effects of visual priors. The eight percussive sounds were presented to the participants through headphones in five random-order sets without any visual stimuli. This phase of the procedure consisted of 40 trials (eight sounds × five presentation sets). The interstimulus interval (ISI) was 1,000–2,000 ms, with an average of 1,500 ms. The uniformity of the surprise levels evoked by the percussive sounds of each video type was confirmed using electroencephalography (EEG) in [9].

Next, the participants watched videos of a clave or a hand drum, which we assumed were familiar instruments for our participants, accompanied by congruent percussive sounds. The videos thus belonged to type AX. The participants watched these videos five times to create expectations of certainty and congruity.

Finally, we conducted the primary experiment in which participants watched videos while undergoing EEG recordings and reporting feelings of surprise subjectively. This experiment contains three sections. The eight videos described in Table 1 were presented to the participants in 20 random-order sets in each section. This phase of the procedure consisted of 480 trials (eight videos × 20 presentation sets × three sections). The ISI was 1,000–2,000 ms, with an average of 1,500 ms. EEG recordings were obtained for each trial. A short break was inserted after the 20th and 40th presentation sets. During the first, 20th, 40th, and final presentation sets, the participants used a four-level Likert scale to report the intensities of their surprise upon listening to the percussive sounds.

TABLE I. COMBINATION OF PERCUSSION INSTRUMENTS AND PERCUSSIVE SOUNDS

|  | *Instrument* | *Congruent sound (X)* | *Incongruent sound (Y)* |
|---|---|---|---|
| Familiar (A) | Clave | Clave (AX) | Bell (AY) |
|  | Hand drum | Hand drum (AX) | Guiro (AY) |
| Unfamiliar (B) | Jawbone | Jawbone (BX) | Vibraphone (BY) |
|  | Slit drum | Slit drum (BX) | Snare (BY) |

## B. EEG

The EEG data were recorded using a portable digital recorder (eego sports, Ant neuro Corporation, Hengelo, Netherlands) and active electrodes. The data were obtained from three midline electrodes positioned at the Fz, Cz, and Pz points as defined by the international 10-20 system with reference to the nose. The data were recorded at a sampling rate of 500 Hz. The time constant was set as 3 s. All electrode impedances were below 60 kΩ. A digital bandpass filter of 0.1–20 Hz was applied.

The ERP waveforms were obtained by averaging data from the period starting at 200 ms before the stimulus onset, which we define as the start of the video in the video stimulus sessions, and ending at 1,500 ms after the stimulus onset. This averaging was performed separately for each participant, stimulus type (i.e., AX, AY, BX, and BY), and electrode site for both the sound-only and video stimuli. For each averaged waveform, the 200-ms period preceding the stimulus onset was defined as the baseline. Any epochs containing EEG signals exceeding ±100 μV were regarded as eye-movement related artefacts and removed automatically. The P300 component was designated as the largest positive peak occurring 250-600 ms after the onset of the percussive sound. The baseline-to-peak P300 amplitudes were measured at the Pz point, which was the dominant electrode site.

Repeated-measures analysis of variance (ANOVA) was applied to the ERP and Likert scale data. To identify the interaction effects on the surprise intensities, we analyzed the P300 amplitude and Likert scale data from the video sessions with a three-way ANOVA in terms of congruity, familiarity, and number of exposures. The statistical significance was defined as p < 0.05 for all statistical tests. We compared the experimental results to the simulation results shown in Figs. 2 and 3.

*C. Experimental Results*

Fig. 5 shows the average P300 amplitude for a number of exposures for each condition of the samples (i.e., familiarity and congruency). The simple primary effect of familiarity was significant (F = 6.06, p = 0.016). The simple primary effect of congruity and that of number of exposures were not significant. However, Fig. 5 suggested that the average P300 amplitude for the unfamiliar instruments decreased larger than that for the familiar instruments under an incongruent sound (AY and BY in Fig. 5).

Fig. 6 shows the average Likert scale surprise rating for each stimulus. The significant interaction effect between uncertainty and number of exposure was not found. The interaction effect of congruity and familiarity was significant (F = 15.21, p < 0.001). The simple primary effect of familiarity was significant at the first, 40th, and 80th exposure for the congruent sound. These results indicate that subjectively rated surprises under the familiar instruments were larger than those under the unfamiliar instruments when the number of exposure is small. The difference of that between the familiar and unfamiliar instruments disappears as the number of exposure increases (AX and BX in Fig. 6). This implies a subjectively rated surprise for the unfamiliar instruments decreased larger with repeated exposure. The primary effect of familiarity for the incongruent sound was insignificant at any number of exposures; however, the score of surprise for unfamiliar instruments tended to converge to a certain value faster than that for familiar instruments (AY and BY in Fig. 6).

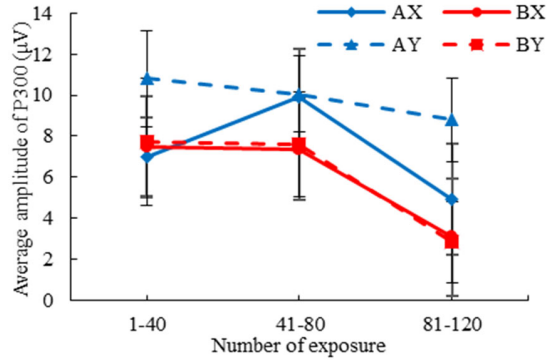

Figure 5. P300 amplitudes evoked by percussive sounds with different congruencies of the instrument shown at every 40 exposures

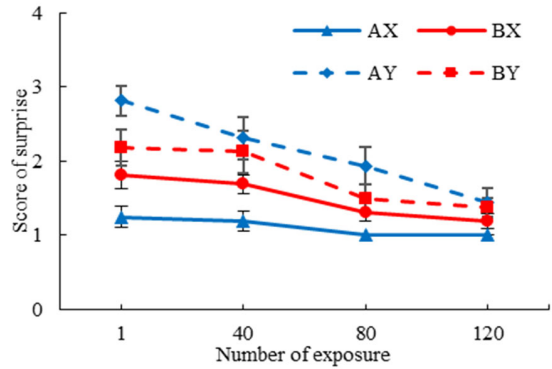

Figure 6. Subjectively reported scores for surprise intensities in response to percussive sounds with different congruencies of the instrument shown at every 40 exposures

## V. DISCUSSION

We assumed that emotional habituation can be represented by a decrement in information gain, which is calculated by the difference of information entropy between the prior expectation and the posterior experience. We described the decrement in information gain as updating the posterior experience to a newly generated expectation and becoming closer to a likelihood function gradually (Bayesian update). We derived the information gain from the repeated exposure of a novel stimulus as a function of three parameters: initial prediction error, initial uncertainty, and noise of sensory stimuli. With the proposed model, we found an interaction effect between the initial prediction error and initial uncertainty on habituation. Finally, we hypothesized that the greater the initial uncertainty, the faster one becomes accustomed to repeatedly exposed novel stimuli.

We conducted an experiment using a set of short videos of percussion instruments accompanied by synthesized percussive sounds. We manipulated the initial prediction error with the congruency of percussive sounds, and the initial uncertainty with the familiarity of instruments shown. We used the average P300 amplitudes and subjectively reported score for surprise intensities as the index of how the participants were surprised by the percussive sounds. Form the experimental result of P300 and subjective score showed the tendency corresponding the hypothesis: the less familiar the object, the faster one becomes accustomed to novel stimuli.

We formalized the acceptable novelty as a function of the initial uncertainty based on Berlyne's theory. We demonstrated that the range of positive feelings shifted toward a more novel experience by repeated novelty exposure. This simulation results supported the experimental results of the previous study where Lévy et al. demonstrated that exposure to a stimulus with a slightly higher novelty than one's preferred level of perceived novelty caused a shift toward more a novel experience level. Meanwhile, the effect of uncertainty on the shift of preference has not yet been validated experimentally.

Previous studies have shown experimentally that the novelty of a stimulus and emotional perception decreased with repeated exposure. We newly formalized habituation as a decrement in information gain through Bayesian update. This model suggests that the ease of habituation to novelty (ease to become accustomed to novelty) differs depending on the initial uncertainty that is affected by contextual factors such as prior information, personal knowledge and experience, and familiarity with an event even if the deviation (prediction error) from the prior expectation is the same. We expect this model to be applicable as a mathematical index to predict one's emotion and to product design that provides a longer-lasting novel impression. In the future, we will validate our valence model by verifying the simulation results with experimental evidence.

## Acknowledgment

This study was supported by KAKEN grant number 18H03318 from the Japan Society for the Promotion of Science. We thank Dr. Kazutaka Ueda of University of Tokyo for supporting ERP measurement in the experiment.